\documentclass[aps,prb,twocolumn,superscriptaddress,english]{revtex4}
\usepackage{amssymb}
\usepackage{graphicx}
\usepackage{amsmath}
\usepackage{soul}
\usepackage{amsthm}
\usepackage{amsfonts}
\usepackage[T1]{fontenc}
\usepackage[latin9]{inputenc}
\usepackage{array}
\usepackage{multirow}
\usepackage{color}
\usepackage{esint}
\usepackage{bm}
\usepackage{color}
\usepackage{bbm}
\usepackage{hyperref}
\usepackage{babel}
\usepackage{titlesec}

\newcommand{\ket}[1]{\left|{#1}\right\rangle}
\newcommand{\bra}[1]{\left\langle{#1}\right|}

\newcommand{\Tr}{{\mathrm{Tr}}}

\begin{document}
	
	\global\long\def\id{\mathbbm{1}}
	\global\long\def\ui{\mathbbm{i}}
	\global\long\def\ud{\mathrm{d}}
\title{Inducing a transition between thermal and many-body localized states and detecting many-body mobility edges through dissipation}
\author{Yutao Hu}
\thanks{These authors contribute equally to this work.}
\affiliation{Shenzhen Institute for Quantum Science and Engineering,
	Southern University of Science and Technology, Shenzhen 518055, China}
\affiliation{International Quantum Academy, Shenzhen 518048, China}
\affiliation{Guangdong Provincial Key Laboratory of Quantum Science and Engineering, Southern University of Science and Technology, Shenzhen 518055, China}
\author{Chao Yang}
\thanks{These authors contribute equally to this work.}
\affiliation{Shenzhen Institute for Quantum Science and Engineering,
	Southern University of Science and Technology, Shenzhen 518055, China}
\affiliation{International Quantum Academy, Shenzhen 518048, China}
\affiliation{Guangdong Provincial Key Laboratory of Quantum Science and Engineering, Southern University of Science and Technology, Shenzhen 518055, China}
\author{Yucheng Wang}
\thanks{Corresponding author: wangyc3@sustech.edu.cn}
\affiliation{Shenzhen Institute for Quantum Science and Engineering,
	Southern University of Science and Technology, Shenzhen 518055, China}
\affiliation{International Quantum Academy, Shenzhen 518048, China}
\affiliation{Guangdong Provincial Key Laboratory of Quantum Science and Engineering, Southern University of Science and Technology, Shenzhen 518055, China}

\begin{abstract}
	The many-body mobility edge (MBME) in energy, which separates thermal states from many-body localization (MBL) states, is a critical yet controversial concept in MBL physics. Here we examine the quasiperiodic $t_1-t_2$ model that features a mobility edge. 
	With the addition of nearest-neighbor interactions, we suggest the potential existence of a MBME. Then we investigate the impact of a type of bond dissipation on the many-body system by calculating the steady-state density matrix and analyzing the transport behavior, and demonstrate that dissipation can cause the system to predominantly occupy either the thermal region or the MBL region, irrespective of the initial state. Finally, we discuss the effects of increasing system size. Our results indicate that dissipation can induce transitions between thermal and MBL states, providing a new approach for experimentally determining the existence of the MBME.
\end{abstract}
\maketitle
\section{Introduction}
Closed quantum many-body systems with disorder or quasiperiodic potentials can exhibit localization~\cite{review1,review2,review3}, wherein the system cannot act as its own heat bath and thus does not reach thermodynamic equilibrium.
Many-body localization (MBL) exhibits several intriguing properties, such as the absence of conductivity (even at finite temperatures)~\cite{review2,Basko2006,Gornyi2005}, specific spectral properties~\cite{Huse2007,Huse2010,Serbyn2016}, entanglement entropy that follows an area law~\cite{review1,review2}, and the slow logarithmic growth of entanglement entropy~\cite{Chiara2006,Prosen2008,Moore2012,Serbyn2013}. In addition to its fundamental theoretical importance, MBL has potential significant applications in quantum information~\cite{Altman2015,Nayak2014,CYin2024}, time crystals~\cite{NormanYao}, and other areas, which has garnered widespread theoretical and experimental attention.
For example, MBL phenomena have been observed in platforms such as cold-atom systems~\cite{IBloch2015,IBloch2016,Islam2015}, circuits with superconducting qubits~\cite{Roushan2017,HHWang2021} and trapped-ion~\cite{JSmith2016} systems.

However, even in these highly controlled implementations, MBL systems are influenced by at least a slight coupling to the environment, such as inelastic scattering from lasers, which may ultimately disrupt localization. Furthermore, considering the observation of MBL in traditional solid-state experiments~\cite{Lake2207} and the applications of MBL phenomena, the impact of dissipation on MBL becomes even more unavoidable. Therefore, the impact of dissipation on MBL has attracted extensive research attention~\cite{JSmith2016,DAHuse2014,Johri2015,DAHuse2015,Hyatt2017,Fischer2016,Levi2016,Medvedyeva,Everest2017,Knap2017,LNWu2019,JRen2020,Gopalakrishnan,Roeck2017,Luitz2017,Richter2024,Denisov2018,Bloch2017,Bloch2019,Brighi2023}.
When coupled to a thermal bath, dissipation can lead to infinite heating in the long-time limit, making MBL typically considered unstable~\cite{DAHuse2014,Johri2015,DAHuse2015,Hyatt2017,Fischer2016,Levi2016,Medvedyeva,Everest2017,Knap2017,LNWu2019,JRen2020,Gopalakrishnan,Roeck2017,Luitz2017}, as observed in cold-atom experiments~\cite{Bloch2017,Bloch2019}. Although dissipation can disrupt MBL, transport in the dissipation-induced delocalized phase is different from that in a typical thermalized phase~\cite{Knap2017}. 

The concept of the mobility edge (ME)~\cite{ME1} is a central idea in the field of localization physics. Similar to how MBL can be seen as the counterpart of Anderson localization~\cite{PWAnderson} in many-body systems, the ME can be extended to many-body systems, suggesting the existence of the many-body mobility edge (MBME)-- a critical energy in the spectrum that separates thermal and MBL eigenstates. Numerous numerical results and some experiments suggest the existence of the MBME~\cite{CYin2024,MBME0,MBME1,MBME2,MBME3,MBME4,MBME5,MBME6,MBME7,MBME8}. However, due to computational size limitations, the existence of a definitive MBME remains uncertain. For instance, Ref.~\cite{noME} argues that local fluctuations in the system with a putative MBME can act as mobile bubbles, inducing global delocalization, and hence, a MBME cannot exist. However, this picture has raised some doubts, as Ref.~\cite{Brighi2020} demonstrates that MBL remains stable in the presence of small bubbles in large dilute systems.
If a MBME exists, could the introduction of dissipation lead to some interesting results? 

Recently, Liu et al. applied a bond-dissipative operator (given by Eq. (\ref{jump}) below) to a system with single-particle MEs and discovered that it can induce a transition between extended and localized states~\cite{BD5}. If such dissipation is applied to systems with MBMEs, could it induce a transition between many-body localized and thermal states? If so, this would mean that we can manipulate the transport properties of these many-body systems. Additionally, it would imply that we could separately measure the properties of many-body localized and thermal states, providing new directions for experimental detection of MBMEs.
In this work, we demonstrate that by introducing such dissipation into a system that, in terms of its computable sizes,  exhibits the MBME, we can drive the system into a steady state, which predominantly consists of either thermal states or many-body localized states, independent of the initial state. Therefore, within the limits of computable finite sizes, we can assert that dissipation can induce transitions between thermal states and MBL states. At the same time, our results also provide a feasible approach for determining the existence of the MBME in simulated systems such as cold atoms.

\section{Models and Results}
\subsection{Quasiperiodic $t_{1}-t_{2}$ model}
We consider the quasiperiodic $t_{1}-t_{2}$ model~\cite{t1t2} with nearest-neighbor (NN) interactions, given by
\begin{eqnarray}
	H &=&-\sum_{j}\left( t_{1}c_{j+1}^{\dag }c_{j}+t_{2}c_{j+2}^{\dag }c_{j}+%
	\text{H.c.}\right) +U\sum_{j}n_{j}n_{j+1}  \nonumber \\
	&&+2\lambda \sum_{j}\cos \left( 2\pi \omega j+\phi \right) n_{j},  \label{H}
\end{eqnarray}%
where $t_{1}$ and $t_{2}$ are the NN and next-NN hopping amplitudes, respectively,
$U$ is the NN interaction strength, $\omega $ is an irrational number, $\lambda$ is the quasiperiodic potential strength, and $\phi $ is a phase offset. Without loss of generality, we take $t_1=1$, $t_2=0.2$, and 
$\omega =\left( \sqrt{5}-1\right) /2$. Unless otherwise stated, we will use open boundary conditions in subsequent computations.

\begin{figure}[t]
	\centerline{\includegraphics[width=1\linewidth]{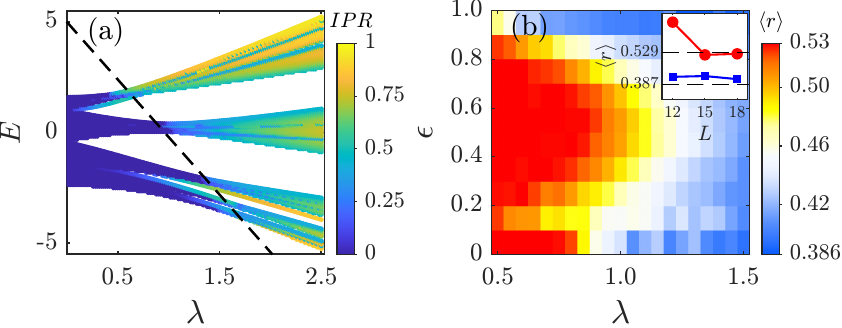}}
	\caption{ (a) IPR of various eigenstates plotted against their corresponding eigenvalues 
		$E$ and the strength of the quasiperiodic potential of the
		$t_{1}-t_{2}$ model. Black dashed line indicates the approximate mobility edge obtained from Eq. (\ref{ME}).
		(b) Variation of $\left\langle r\right\rangle$ with energy density ($\epsilon$) and quasiperiodic potential strength $\lambda$. Here $t_{1}=1$ and $t_{2}=0.2$, $U=0.9$, 
		the system size $L=18$, the number of particles $N=6$, and the number of samples is $20$, with each sample corresponding to an initial phase $\phi$. Inset: Variation of $\left\langle r\right\rangle$
		with system size at fixed $\lambda=0.8$ and one-third filling, for the highest (blue) and lowest (red) energy window.}
	\label{fig1}
\end{figure}

In the non-interacting limit ($U\rightarrow 0$), this model displays a single-particle ME, as shown in Fig. \ref{fig1}(a), where we present the inverse participation ratio (IPR) of each eigenstate, which, for an arbitrary 
$m$-th eigenstate $|\psi_m\rangle=\sum_{j}^L\psi_{m,j}c_{j}^{\dagger}|\varnothing\rangle$ with $L$ being the system size, is defined as $IPR=\sum_{j=1}^{L}\left\vert \psi _{m,j}\right\vert ^{4}$. It is known that for extended and localized states, the IPR tends to $0$ or a finite nonzero value, respectively. From Fig. \ref{fig1}(a), we see that the extended and localized states are divided by one ME. In general, the $t_{1}-t_{2}$ model does not have an analytical expression for the ME. However, when $t_1/t_2$ is sufficiently large, the NN and next-NN hopping terms can be approximated by an exponential hopping with strength $t=t_1e^{p}$ with $p=\ln(t_1/t_2)$~\cite{Biddle1,Biddle2}. The latter has an exact expression for the ME, and based on this, we can obtain an approximate expression for the ME of the $t_{1}-t_{2}$ model in this case: 
\begin{equation}\label{ME}
	E_c=\frac{-\lambda(t_1^2+t_2^2)+t_1^3}{t_1t_2},
\end{equation}
as shown by the dashed line in Fig. \ref{fig1}(a).

When the NN interaction is included, the single-particle ME may lead to the emergence of a MBME. To characterize the localized and thermal properties of this system, we consider the ratio of adjacent energy gaps~\cite{Huse2007,Huse2010,Serbyn2016} as $r_{i}=\frac{\min \left( \delta _{i+1},\delta _{i}\right) }{\max \left(\delta _{i+1},\delta _{i}\right) }$, where $\delta _{i}=E_{i}-E_{i+1}$ represents the energy spacing with the eigenvalues $E_i$ listed in ascending order.  The range of $r$ is from $0$ to $1$, making it more convenient to use $r$ instead of $\delta$ for analysis. $r$ reflects the uniformity of the level spacing distribution. When the level distribution is perfectly uniform, meaning all level spacings are equal, 
$r=1$. Therefore, a larger $r$ indicates a more uniform level distribution. For the system in the MBL region, 
the absence of correlation between energy levels results in Poissonian level statistics, with $\left\langle r\right\rangle \approx 0.39$.
In the thermal region, its level statistics follow the Wigner-Dyson distribution, where level repulsion leads to a more uniform distribution, and the average $\left\langle r\right\rangle$ converges to $0.53$.  We rescale the many-body spectrum as $\epsilon _{i}=\left(
E_{i}-E_{g}\right) /\left( E_{\max }-E_{g}\right) $, where $E_{g}\left(
E_{\max }\right) $ is the eigenenergy of the ground (highest excited) state.
Based on this, we divide the eigenvalues into $10$ different energy windows, each containing an equal number of eigenvalues,  and average samples and all gaps in each window to obtain an $\left\langle r\right\rangle$, as shown in Fig. \ref{fig1}(b).
We observe that as the strength of the quasiperiodic potential $\lambda$ increases, the states in the uppermost window become localized first, similar to the single-particle case. Due to the presence of interactions, the mixing of extended and localized states in the middle region of the spectrum disrupts the original position of the ME. However, the mixing of states within the localized region at the top (or the extended region at the bottom) of the spectrum still preserves their localized (or extended) nature, forming many-body localized (or thermalized) states. In the inset of Fig. \ref{fig1}(b), we fix $\lambda=0.8$. The red and blue lines correspond to the $\left\langle r\right\rangle$ of the lowest energy window and the highest energy window, respectively, which converge to $0.53$ and $0.39$~\cite{explainB}.  This indicates that the states in the highest window become localized while the states in the lowest window remain thermalized, suggesting the existence of a MBME. In the following discussion, we fix $\lambda=0.8$ and one-third filling.

\subsection{Dissipation-induced transition between thermal and many-body localized states}
Next, we introduce the dissipation affecting a pair of neighboring sites $j$ and $j+1$, described by~\cite{BD1,BD2,BD3,BD4,BD5,BD6,BD7,BD8,BD9} 
\begin{equation}
	O_{j}=(c_{j}^{\dag }+ac_{j+1}^{\dag })(c_{j}-ac_{j+1}),  \label{jump}
\end{equation}%
where $a=\pm 1$, and $j=1,2,\ldots,L-1$. This form of dissipation can be implemented using cold atoms in optical superlattices~\cite{BD1,BD2,BD3,BD4} or through arrays of superconducting microwave resonators~\cite{BD6}. This bond dissipation maintains the particle number but modifies the relative phase between adjacent lattice sites. It synchronizes them from an in-phase (out-of-phase) mode to an out-of-phase (in-phase) mode when $a$ is set to $-1$ ($1$), enabling control over the steady state's position within the energy spectrum~\cite{BD5,BD7}.

\begin{figure}[tbph]
	\centerline{\includegraphics[width=1\linewidth]{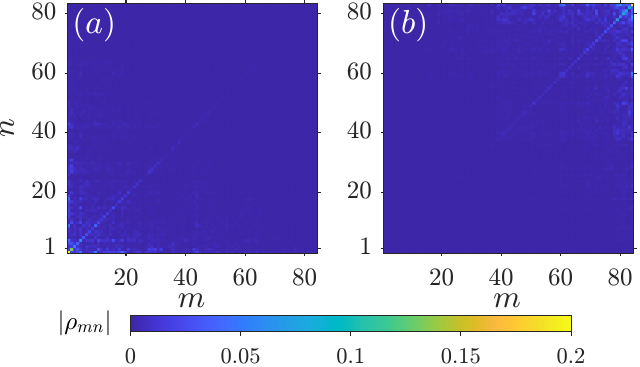}}
	\caption{ Absolute value of steady-state density matrix elements in the
		Hamiltonian eigenstate basis with (a) $a=1$ and (b) $a=-1$. Here we take $L=9$, $N=3$, $U=0.9$, $\lambda =0.8$, and $\Gamma =1$. }
	\label{fig2}
\end{figure}

The dissipative dynamics of density matrix $\rho $ is represented by the Lindblad master eqaution~\cite{GLindblad,HPBreuer}
\begin{eqnarray}
	\frac{\ud\rho(t)}{\ud{}t} &=& \mathcal{L}\left[ \rho \left( t\right) \right] =-i\left[ H,\rho \left(
	t\right) \right]   \nonumber \\
	&&+\Gamma\sum_{j}\left( O_{j}\rho O_{j}^{\dag }-\frac{1}{2}\left\{
	O_{j}^{\dag }O_{j},\rho \right\} \right) ,  \label{L}
\end{eqnarray}%
where $\mathcal{L}$ is the Lindbladian superoperator, and the strength of jump operators is set to 
$\Gamma$, which is independent of the lattice points. Our results are essentially independent of the value of 
$\Gamma$, and without loss of generality, we set $\Gamma=1$. Here, $\mathcal{L}$ is time-independent, so we can express $\rho(t)=e^{\mathcal{L}t}\rho(0)$. The steady state is defined as $\rho_{s}=\rho(t\rightarrow\infty)$, which corresponds to the eigenstate of the Lindbladian $\mathcal{L}$ with the zero eigenvalue, i.e., $\mathcal{L}[\rho_{s}] = 0$. 
We analyze the properties of the steady-state density matrix $\rho_{s}$ in the eigenbasis of the many-body Hamiltonian $H$, that is, $\rho_{nm}=\langle \psi_n|\rho_{s}|\psi_m\rangle$, where $|\psi_n\rangle$ and $|\psi_m\rangle$ denote the eigenstates of $H$. Fig. \ref{fig2} illustrates that the system's steady state primarily occupies the low-energy thermal region when $a=1$ [Fig. \ref{fig2}(a)], whereas it predominantly occupies the high-energy MBL region when $a=-1$ [Fig. \ref{fig2}(b)]. 
This means that by adjusting dissipation, we can control whether the steady state of this system predominantly resides in the thermalization region or in the MBL region.

When dissipation is present, we note that even if the steady state of the system resides in a localized region under the basis of the Hamiltonian, the system's dynamics are not necessarily localized. This is because the bond dissipation described by Eq. (\ref{jump}) in our study expands to include terms related to dephasing dissipation $c^{\dagger}_jc_j$, which which can destroy localization by disrupting coherence. Therefore, we also need to remove dissipation after the system reaches a steady state. Dephasing dissipation destroys localization, but once dissipation is removed after the system reaches a steady state, coherence is restored, and the system recovers its localized state. However, the dissipation we introduced modifies the MBL or thermal states through a different mechanism, namely by selecting certain states. Once the system reaches a steady state, this state selection has already been completed. Upon removing dissipation, the effects induced by the bond dissipation do not vanish. This can be observed through the evolution of the density matrix after dissipation is removed. After abruptly removing the dissipation, we observe
the evolution of the density matrix: $\rho(t)=e^{iHt}\rho e^{-iHt}=\sum_{mn}e^{i(E_m-E_n)t}\rho_{mn}|\psi_m\rangle\langle \psi_n|$, where $E_m$ and $E_n$ are the eigenvalues corresponding to the many-body eigenstates $|\psi_m\rangle$ and $|\psi_n\rangle$.
We observe that the diagonal elements of $\rho$ (where $E_m=E_n$) do not change with time, while the off-diagonal elements fluctuate over time. However, with longer observation times, we need to examine the average effects of the dynamics over this interval, leading to the vanishing of the effects caused by the off-diagonal elements~\cite{explainoff}.
Therefore, when the steady state primarily occupies the localized (thermal) region, removing dissipation reveals behavior characteristic of the localized (thermal) region. A superposition of localized (thermal) states remains localized (thermal). Based on this observation, we can conclude that dissipation serves as an intermediate process capable of inducing a transition between thermalization and MBL states.

Changes between thermalization and MBL induced by dissipation will inevitably cause changes in transport properties~\cite{review2,Basko2006,Gornyi2005,Varma2016}, and these transport properties can be probed by experimental platforms such as ultracold atoms~\cite{transportExp1,transportExp2,transportExp3,transportExp4,transportExp5}.  By using these properties to distinguish between thermalization and MBL behaviors, we can experimentally verify that dissipation can induce transitions between thermal and localized states. 
We assume the system reaches a steady state at the time $t_0$, after which dissipation is removed. We mainly focus on the transport properties after this point. The system then returns to a non-equilibrium state, and we study its response to a probed electric field. The change in current $\delta I$, for a weak probed electric field, can be derived from linear response theory.
\begin{equation}
	\langle\delta I(t)\rangle=\int_{t_0}^{t}L\sigma(t,t^{'})E(t^{'})dt^{'},
\end{equation}
where $\sigma(t,t^{'})$ is the non-equilibrium conductivity, which depends only on the time difference $t-t^{'}$ for an equilibrium state~\cite{transport1}. Here we consider a delta probed field $E(t)=E\delta(t-t_0)$ such that $\langle\delta I(t)\rangle=LE\sigma(t,t_0)$. For convenience, we set $e=\hbar=E=1$. So the conductivity in a finite size system can be written as \cite{transport2}
\begin{equation}
	\sigma(t,t^{'})=\frac{i}{N}\theta(t-t^{'}) Tr[\rho(t_0)[I(t),B(t^{'})]],
\end{equation}
where the position operator $B=\sum_{j}jc_{j}^{\dagger}c_{j}$ and the current operator $I=\frac{dB}{dt}=-i[B,H]$.
The current change in the thermal region is expected to show large fluctuations because it is significantly affected by the electric field, whereas in the MBL state, the current change should exhibit smaller fluctuations due to minimal influence from the electric field~(see Appendix C). In Fig. \ref{fig3}, using the same parameters as in Fig. \ref{fig2}, we show $\delta I$ for the cases $a=1$ (blue line) and $a=-1$ (orange line). When $a=1$, as discussed earlier, the steady state primarily occupies the thermalization region near the ground state. After removing dissipation, we observe that the current change consistently exhibits larger oscillations. In contrast, when $a=-1$ (the orange line in Fig. \ref{fig3}), the system's steady state mainly occupies the many-body localized region near the highest state. After removing dissipation, we see that the current change oscillates with a smaller amplitude and evolves to near zero at a faster rate, exhibiting relatively localized properties.
\begin{figure}[tbph]
	\centerline{\includegraphics[width=1\linewidth]{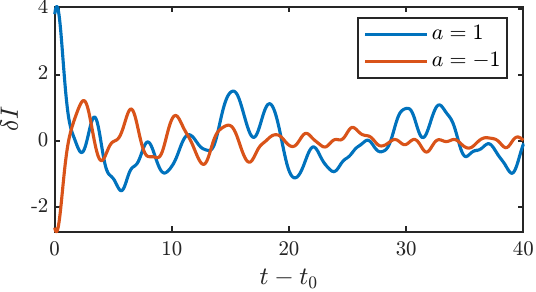}}
	\caption{ The conductance of the steady-state density matrix after removing dissipation varies with time. The red line corresponds to the case of $a=-1$, where the steady state mainly occupies the localized region near the highest excited state. In contrast, the blue line corresponds to $a=1$, where the steady state mainly occupies the thermalized region near the ground state. Other parameters are the same as in Fig. 2. }
	\label{fig3}
\end{figure}

\begin{figure}[t]
	\centerline{\includegraphics[width=1\linewidth]{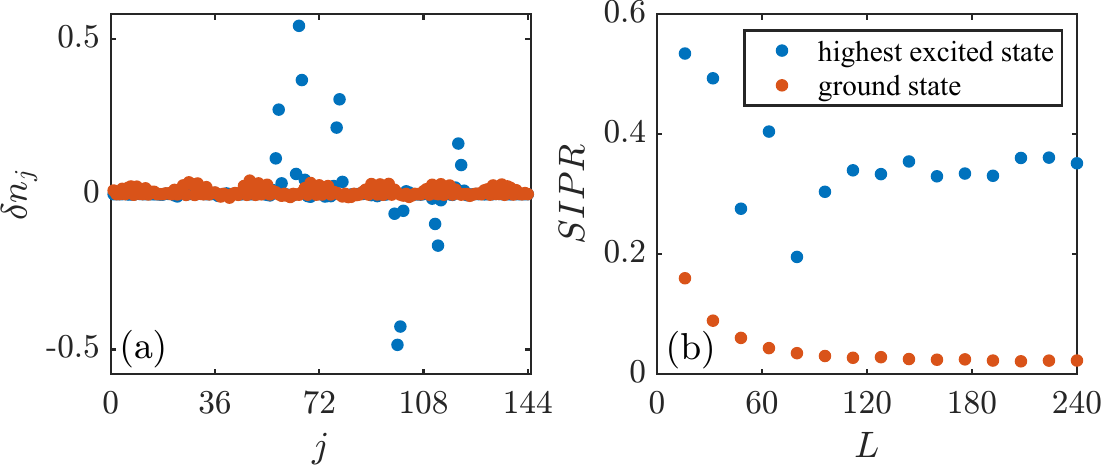}}
	\caption{ (a) the distribution of $\delta n$ at each lattice site after adding a particle to a system with 
		$L=144$ and $N=32$. (b) The variation of SIPR with system size. The orange and blue dots represent the ground state and the highest excited state, respectively. Here, we always retain $n_s = 40$ states when truncating the reduced density matrix and perform $20$ sweeps in the DMRG calculation.} 
	\label{fig4}
\end{figure}
\subsection{Discussion on the effects of system size}
As discussed earlier, in small systems, there exists a MBME, and dissipation can induce transitions between thermal and MBL states. However, when the system size increases, the existence of a MBME becomes a controversial issue. If there is no MBME, then the dissipation-induced transitions we discussed here would not exist. Inspired by recent experiments~\cite{DeteME1,DeteME2} that probe the ME by measuring the localized and extended properties of the ground state and the highest excited state, we use the density matrix renormalization group (DMRG) method to study the thermalization and localization properties of the ground state and the highest excited state in large systems. To characterize the localization properties of these two states, we consider the particle insertion response, defined as the difference in distribution after introducing an additional particle into the many-body system, i.e., $\delta n_{j}=\rho _{N+1}\left( j\right) -\rho _{N}\left( j\right)$,
where $\rho _{N}\left( j\right) =\langle \psi^N _{g(e)}|n_{j}|\psi^N
_{g(e)}\rangle $ and $|\psi^N _{g(e)}\rangle $ is the ground (highest excited) state of the system with $N$
particles. Such a quantity has been used to characterize the topologically localized edge excitations~\cite{Hu1,Hu2} and the thermal-MBL transition of the ground state in many-body systems~\cite{WangEPJB}. Similarly, we can also define the IPR of single-particle (SIPR) density change as
\begin{equation}
	SIPR=\frac{\sum_{j}\delta n_j^2}{\sum_{j}|\delta n_j|}.
\end{equation}
If the ground state or the highest excited state is thermal, this added particle density profile should be extended, and therefore the SIPR should approach $0$. However, if the ground state or the highest excited state is many-body localized, the added particle's effect on the system does not propagate to distant regions, meaning this effect is localized. In this case, the added particle density profile should be localized, and the corresponding SIPR will not approach $0$ but instead settle at a finite value. We add a particle to a system with size $L=144$ and particle number 
$N=32$, and show the distribution of $\delta n$ at each lattice site in Fig. \ref{fig4}(a). We see that for the ground state (the orange dots), the single particle excitation is distributed quite uniformly across different lattice sites, indicating an extended property. In contrast, for the highest excited state (the blue dots), $\delta n_{j}$ is primarily concentrated on a few lattice sites, showing localized behavior. Fig. \ref{fig4}(b) shows the variation of the SIPR with system size. We observe that for the ground state, the SIPR approaches $0$ as the size increases, while for the highest excited state, the SIPR eventually stabilizes at a certain value and does not decrease with increasing size. These results correspond to those in Fig. \ref{fig1}, indicating a significant difference in localization properties between the ground state and the highest excited state in larger systems. We note that the ground state and the highest excited state exhibit different localization properties, which, apart from the possible existence of a MBME, could also arise from boundary effects in finite-size systems, topological effects (such as the presence of topological boundary states or defects), electron-phonon coupling, or irregular band structures, among others. However, in the context of the system we study-where a ME is present in both single-particle and small-size many-body systems-the most likely explanation lies in the existence of a MBME. This interpretation is consistent with recent experiments~\cite{DeteME1,DeteME2} that observed distinct localization behaviors of the ground and highest excited states as evidence for the existence of a ME. While we do not claim to numerically prove the existence of a MBME, this approach may offer useful insights into addressing finite-size effects in its investigation.

In the DMRG calculation, we always retain $40$ states when truncating the reduced density matrix. Changing the number of retained states does not significantly affect the $\delta n_j$ and SIPR for the highest excited state, as its entanglement satisfies the area law, and DMRG can provide accurate results. However, for the ground state, changing the number of retained states during the calculation causes noticeable changes in $\delta n_j$, which indicates that the DMRG results for the thermal state are not precise. Nevertheless, $\delta n_j$ remains delocalized, and the SIPR does not change significantly with the number of retained states (see Appendix D). Therefore, while the DMRG results for the ground state are not accurate, they are sufficient to demonstrate its delocalized properties.

\subsection{Discussion on detecting the MBME}
Our results suggest that the MBME might exist in large systems. In this case, the transition between thermal and MBL states induced by dissipation should also be present in large systems. This provides a potential method for experimentally determining whether the MBME exists. Although the MBME has been observed in some systems, such as superconducting qubit circuits~\cite{HHWang2021}, the system sizes used in experiments are still not large enough. Cold atom systems can simulate larger systems, but measuring the MBME in the system presents significant challenges~\cite{BlochME}. Based on our results, by adjusting dissipation, we can selectively place the system in either the thermal region or the MBL region, facilitating measurements. This makes it possible to detect the MBME in cold atom experiments.

\section{Conclusion}
We have investigated the impact of a type of bond dissipation on the quasiperiodic $t_1-t_2$ model with NN interactions, which possesses a mobility edge in the absence of interactions. By calculating the level spacing distribution, we observed
signatures consistent with the presence of a MBME. We further analyzed the distribution of the steady-state density matrix and revealed that dissipation can drive the many-body system into specific states primarily located in either the thermal region or the MBL region, regardless of the initial states. 
Thus, dissipation can be used to induce transitions between thermal and MBL states, allowing for the control of particle transport behaviors, as demonstrated by our analysis of the system's transport properties after removing dissipation. We note that bond dissipation only needs to act for a short period of time to drive the system into a steady state, after which it can be removed. Moreover, even substantial fluctuations in the strength of the bond dissipation do not alter the steady-state results. Therefore, the physical effects induced by this dissipation are highly stable.
Finally, we used DMRG to study single-particle excitations in the ground state and highest excited state for large system sizes, demonstrating their different localization behaviors, which implies that MBME may exist at large sizes. Therefore, the dissipation-induced thermal-MBL transition should also exist at large sizes, providing a novel method for experimentally identifying the presence of the MBME.

\begin{acknowledgments}
This work is supported by National Key R\&D Program of China under Grant No.2022YFA1405800, the National Natural Science
Foundation of China (Grant No.12104205), the Key-Area Research and Development Program of Guangdong Province (Grant No. 2018B030326001), Guangdong Provincial Key Laboratory (Grant No.2019B121203002).	
\end{acknowledgments}	

\appendix
\section {Compute the steady state}
To numerically compute the steady state, we represent the density matrix $\rho$ as a vector and the Liouvillian superoperator
$\mathcal{L}$ as a matrix. In the computational basis $\{\ket{k}\}$, the density matrix can be expressed as
$\rho=\sum_{j,k}\rho_{jk}\ket{j}\bra{k}$, which is then mapped to a vector representation as:
\begin{equation}
	\ket{\rho} = \sum_{j,k}\rho_{jk}\ket{j}\otimes\ket{k}.
	\label{eq_rho}
\end{equation}
Consequently, the Liouvillian superoperator  $\mathcal{L}$ can be represented as a matrix:
\begin{eqnarray}
	\mathcal{L} 
	&=&\sum_j\left[{O}_j\otimes{O}_j^*
	-\frac{1}{2}\left({O}_j^\dagger{O}_j\otimes\mathbbm{1}
	-\mathbbm{1}\otimes{O}_j^\mathrm{T}{O}_j^*\right)\right]
	\nonumber\\
	&&-i\left({H}\otimes\mathbbm{1}-\mathbbm{1}\otimes{H}^\mathrm{T}\right).
	\label{eq_Lmatrix}
\end{eqnarray}
We diagonalize the matrix representation of $\mathcal{L}$ to obtain its eigenvalues and eigenvectors $\ket{\rho}$. The eigenvalue of $0$ corresponds to the steady state $\ket{\rho_s}$. Using the mapping defined in \eqref{eq_rho}, we can reconstruct the steady-state density matrix $\rho_s$. For a Hamiltonian with a Hilbert space dimension $D_H$, the Liouvillian
$\mathcal{L}$, as described in Eq. \eqref{eq_Lmatrix}, is a matrix of size $D_H^2\times{}D_H^2$. Therefore, for the many-body system with dissipation, we only compute up to $L=9$ in the main text.

\section{Average IPR for different energy windows}
\begin{figure}[h]
	\centerline{\includegraphics[width=0.8\linewidth]{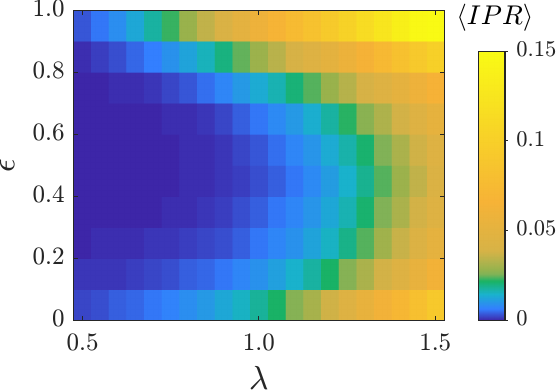}}
	\caption{Variation of $\left\langle {\rm IPR}\right\rangle$ as a function of energy density ($\epsilon$) and quasiperiodic potential strength $\lambda$, using the same parameters as in Fig. 1(b) of the main text.}
	\label{figA1}
\end{figure}

The localization properties of different energy windows shown in Fig. 1(b) of the main text can also be characterized using the inverse participation ratio (IPR), which is defined as ${\rm IPR}=\sum_j|\phi_j|^4$, where $\phi_j$ is the coefficient of the eigenstate $|\psi_n\rangle$ in the computational basis $\{|j\rangle\}$, expressed as $\phi_j=\langle j|\psi_n\rangle$. As the size of the Hilbert space increases, the ${\rm IPR}$ for thermal states tends to zero, while for MBL states, the ${\rm IPR}$ remains a finite nonzero value. Fig. \ref{figA1} shows the average ${\rm IPR}$ within each energy window. Similar to the results of the level statistics in Fig. 1 of the main text, as the quasiperiodic potential strength $\lambda$ increases, the states in the highest energy window localize first. 

\section {Current Changes After Introducing an Electric Field in the Interacting AA Model}
In this section, we take the interacting Aubry-Andr\'{e} (AA) model ($H =-\sum_{j}\left( t_{1}c_{j+1}^{\dag }c_{j}+\text{H.c.}\right) +U\sum_{j}n_{j}n_{j+1}+2\lambda \sum_{j}\cos \left( 2\pi \omega j+\phi \right) n_{j}$, i.e., with next-nearest-neighbor hopping strength $t_2=0$ in the Hamiltonian of Eq. (1) of the main text) as an example to illustrate the different current changes between thermal and many-body localized states after introducing an electric field. We still fix one-third filling.
We consider only the ground state and select two extreme values of the quasiperiodic potential: 
$\lambda=0.1$, representing the thermal state property, and 
$\lambda=10$, representing the MBL state property.

\begin{figure}[h]
	\centerline{\includegraphics[width=0.9\linewidth]{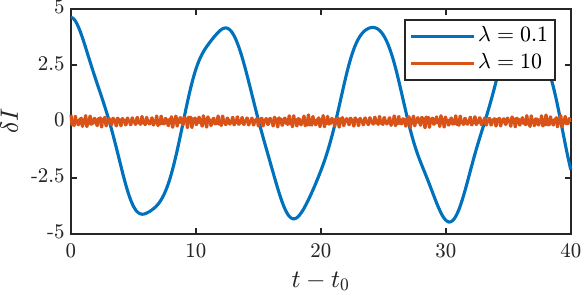}}
	\caption{The current change $\delta I$ of the ground state for the many-body AA model for different quasiperiodic potential strengths: blue line for $\lambda=0.1$ and orange line for $\lambda=10$. Other parameters are $L=9$, $N=3$, $t_1=1$, $t_2=0$, and $U=1$.}
	\label{figB1}
\end{figure}

As shown in Fig. \ref{figB1}, the current change in the thermal state has a large amplitude and a long oscillation period, similar to the case of $a=1$ shown in Fig. 3 of the main text, while the MBL state exhibits a small amplitude around $0$ and fast oscillations, similar to the case of $a=-1$ shown in Fig. 3 of the main text. This can be understood as the current in the thermal state being significantly affected by the electric field, whereas the MBL state is minimally influenced by the electric field.

\section{The Impact of the Number of Retained States on SIPR in DMRG Calculations}
\begin{figure}[h]
	\centerline{\includegraphics[width=0.7\linewidth]{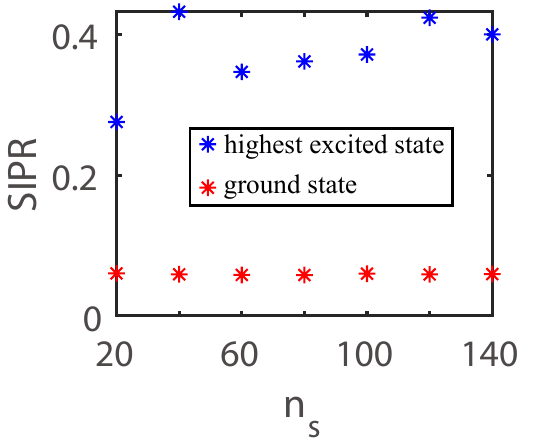}}
	\caption{SIPR as a function of $n_s$ with $L=48$ for the ground state (red dots) and the highest excited state (blue dots). We sweep $20$ cycles in finite size DMRG.}
	\label{figC1}
\end{figure}

For thermal states, their entanglement does not follow the area law. Therefore, it is necessary to clarify that the DMRG calculations used in the main text are valid. As mentioned in the main text, although the $\delta n_j$ values given by DMRG may change, their delocalized nature remains unchanged. In Fig. \ref{figC1}, we fix $L=48$ and a one-third filling, showing how the SIPR of the ground state and the highest excited state varies with the number of retained states $n_s$ in truncating the reduced density matrix. We observe that the SIPR corresponding to the added particle density profile of the thermal state (red data points) does not change with $n_s$. This indicates that the SIPR for the thermal state obtained from our DMRG calculations is valid, and using SIPR to reflect localized and delocalized properties is feasible.

\begin{thebibliography}{99}
\bibitem{explainB} We observe that at $L=12$, $\left\langle r\right\rangle$ in the lowest energy window is significantly greater than $0.53$, and this discrepancy cannot be improved by increasing the sample size. Therefore, we believe the data at this size does not accurately reflect the level distribution. This issue may arise from the small number of levels in each energy window at this system size. At $L=12$ with one-third filling, the number of levels in each energy window is less than $50$, which can lead to statistical issues when analyzing level spacing distributions.
\end{thebibliography}
\end{document}